\documentclass[prb,showpacs,twocolumn,superscriptaddress]{revtex4-1}
\usepackage{graphicx,hyphenat}
\usepackage{color}

\begin{document}
	
\title{High-field magnetization and magnetic phase diagram of $\alpha$-Cu$_{2}$V$_{2}$O$_{7}$}
	
\author{G. Gitgeatpong}
\affiliation{Department of Physics, Faculty of Science, Mahidol University, Bangkok 10400, Thailand}
\affiliation{ThEP, Commission of Higher Education, Bangkok, 10400, Thailand}
\affiliation{Physics Program, Faculty of Science and Technology, Phranakhon Rajabhat University, Bangkok 10220, Thailand}

\author{M. Suewattana}
\affiliation{Department of Physics, Faculty of Science, Mahidol University, Bangkok 10400, Thailand}

\author{Shiwei Zhang}
\affiliation{Department of Physics, College of William and Mary, Williamsburg, Virginia 23187-8795, USA}

\author{A. Miyake}
\affiliation{The Institute for Solid State Physics, University of Tokyo, Kashiwa 277-8581, Japan}
	
\author{M. Tokunaga}
\affiliation{The Institute for Solid State Physics, University of Tokyo, Kashiwa 277-8581, Japan}

\author{P.~Chanlert}
\affiliation{Department of Physics, Tokyo Institute of Technology, Meguro-ku, Tokyo 152-8551, Japan}

\author{N. Kurita}
\affiliation{Department of Physics, Tokyo Institute of Technology, Meguro-ku, Tokyo 152-8551, Japan}

\author{H. Tanaka}
\affiliation{Department of Physics, Tokyo Institute of Technology, Meguro-ku, Tokyo 152-8551, Japan}

\author{T. J. Sato}
\affiliation{IMRAM, Tohoku University, Sendai, Miyagi 980-8577, Japan}
		
\author{Y. Zhao}
\affiliation{Department of Materials Science and Engineering, University of Maryland, College Park, Maryland 20742, USA}
\affiliation{NIST Center for Neutron Research, National Institute of Standards and Technology, Gaithersburg, Maryland 20899, USA}

\author{K. Matan}
\email[]{kittiwit.mat@mahidol.ac.th}
\affiliation{Department of Physics, Faculty of Science, Mahidol University, Bangkok 10400, Thailand}
\affiliation{ThEP, Commission of Higher Education, Bangkok, 10400, Thailand}
	
\date{\today}
	
\begin{abstract}
High-field magnetization of the spin-$1/2$ antiferromagnet $\alpha$-Cu$_2$V$_2$O$_7$ was measured in pulsed magnetic fields of up to 56 T in order to study its magnetic phase diagram. When the field was applied along the easy axis (the $a$-axis), two distinct transitions were observed at $H_{c1}=6.5$~T and $H_{c2}=18.0$~T. The former is a spin-flop transition typical for a collinear antiferromagnet and the latter is believed to be a spin-flip transition of canted moments.  The canted moments, which are induced by the Dzyaloshinskii-Moriya interactions, anti-align for $H_{c1}<H<H_{c2}$ due to the anisotropic exchange interaction that favors the antiferromagnetic arrangement along the $a$-axis.  Above $H_{c2}$, the Zeeman energy of the applied field overcomes the antiferromagnetic anisotropic interaction and the canted moments are aligned along the field direction.  Density functional theory was employed to compute the exchange interactions, which were used as inputs for quantum Monte Carlo calculations and then further refined by fitting to the magnetic susceptibility data.  Contrary to our previous report in Phys. Rev. B {\bf 92}, 024423, the dominant exchange interaction is between the third nearest-neighbor spins, which form zigzag spin-chains  that are coupled with one another through an intertwining network of the nonnegligible nearest and second nearest-neighbor interactions. In addition, elastic neutron scattering under the applied magnetic fields of up to 10 T reveals the incommensurate helical spin structure in the spin-flop state.
\end{abstract}
	
	
\maketitle 

\section{Introduction}

A spin-flop transition in collinear antiferromagnetic systems can be observed when a magnetic field is applied parallel to the easy axis of the antiferromagnet. The strength of the applied magnetic field that forces the spins to flop depends on exchange interactions in the systems. The spin-flop transition, if present, causes the spins to reorient themselves perpendicular to the applied magnetic field in order to compromise the exchange-interaction energy with the Zeeman energy. This phemomenon was predicted eighty years ago~\cite{neel} and has been observed in several compounds~\cite{flop1, flop2, flop3, flop4}. Generally, the spin-flop transition can be observed as a single transition with a sudden increase of magnetization $M$ at a critical field $H_{c}$ as well as the change of magnetic susceptibility defined by the slope of the $M-H$ curve below and above $H_c$. However, there are a few cases in which two successive magnetic phase transitions are observed, for example, in the quasi-one-dimensional BaCu$_2$Si$_2$O$_7$ system~\cite{bcso1, bcso2, bcso3}, of which the underlying mechanism is still unresolved. In this article, we report on the two-stage spin reorientation in $\alpha$-Cu$_{2}$V$_{2}$O$_{7}$ using high field magnetization measurements on single crystal samples. Despite a single spin-flop transition being observed in its cousin phase $\beta$-Cu$_{2}$V$_{2}$O$_{7}$\cite{bcvo} or other antiferromagnetic systems~\cite{}, we instead found two successive jumps in the magnetization of $\alpha$-Cu$_{2}$V$_{2}$O$_{7}$ similar to those observed in BaCu$_2$Si$_2$O$_7$. 

\begin{figure}
	\begin{center}
		\includegraphics[width=\columnwidth]{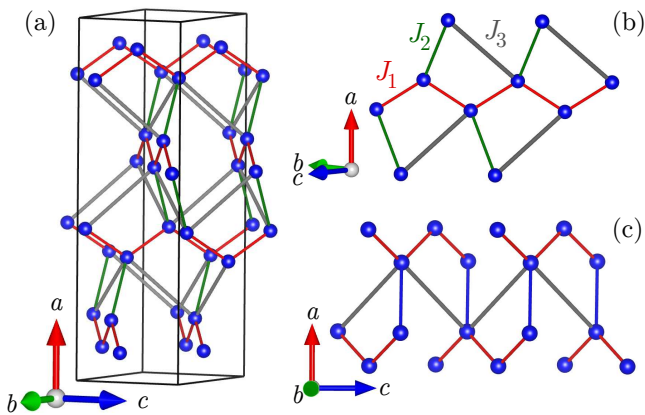}
		\caption{(Color online) Diagrams showing the network of Cu$^{2+}$ ions in $\alpha$-Cu$_{2}$V$_{2}$O$_{7}$. (a) The nearest-, second-nearest, and third-nearest neighbor interactions, $J_1$, $J_2$, and $J_3$, are represented by red, green, and grey lines, respectively. (b) The nearest-neighbor interaction $J_1$ forms zigzag chains which run along the [011] and [01$\bar 1$] directions.  (c) The third nearest-neighbor interaction forms zigzag chain along the $c$-axis.}\label{fig1}
	\end{center}
\end{figure}

$\alpha$-Cu$_{2}$V$_{2}$O$_{7}$ crystallizes in the orthorhombic system ({\sl Fdd2}) with $a$ = 20.645(2)~\AA, $b$ = 8.383(7)~\AA, and $c$ = 6.442(1)~\AA\cite{Calvo1975, Robinson1987}. Below $T_N=33.4$~K, the system undergoes a paramagnetic to antiferromagnetic transition.  In the ordered state, $S=1/2$ Cu$^{2+}$ spins align antiparallel along the crystallographic $a$-axis with their nearest neighbors~\cite{cvo1, cvo2}. The magnetization and powder neutron scattering studies suggest small spin canting along the $c$-axis\cite{cvo1, cvo2} as a result of the antisymmetric Dzyaloshinskii-Moriya (DM) interaction. The exchange interactions in $\alpha$-Cu$_{2}$V$_{2}$O$_{7}$ are, to date, still open to debate. Our previous analysis using quantum Monte Carlo (QMC) simulation~\cite{cvo1} showed two possible models with different values of the nearest-neighbor interaction $J_1$ and second nearest-neighbor interaction $J_2$ that can be equally used to describe the broad maximum observed in the magnetic susceptibility data. On the other hand, density functional theory (DFT) calculations by Sannigrahi {\it et al.}\cite{cvo3} revealed the dominant third nearest-neighbor antiferromagnetic interaction $J_3$ (see Fig.~\ref{fig1} for the diagram). The latest study on a powder sample using inelastic neutron scattering also supports the leading $J_3$ model~\cite{cvo4}.  Both DFT and powder inelastic neutron scattering studies qualitatively suggest that the antiferromagnetic third nearest-neighbor interaction $J_3$ forming zigzag chains along the $c$-axis [Fig.~\ref{fig1}(c)] via a complex Cu--O--V--O--Cu pathway (through the VO$_4$ tetrahedra) is non-negligible and possibly the strongest of the exchange interactions. In addition, the interconnection between electricity and magnetism in $\alpha$-Cu$_{2}$V$_{2}$O$_{7}$ has been studied to reveal its magnetoelectric properties~\cite{cvo3, cvo2}, which might find useful applications.  This variety of interesting phenomena and inconclusive understanding of the nature of the exchange interactions in $\alpha$-Cu$_{2}$V$_{2}$O$_{7}$ have led us to this more detailed investigation of the magnetic properties of the system.

This paper presents a study of the magnetic properties of single-crystal $\alpha$-Cu$_{2}$V$_{2}$O$_{7}$. The experimental details are described in Sec.~\ref{experiment}.  In Sec.~\ref{mpms}, we discuss the magnetization measurements at low field. In Sec.~\ref{dft}, the DFT calculation and QMC simulation are discussed and compared to the low-field magnetic susceptibility data. In Sec.~\ref{highfield}, we investigate the magnetic phase transitions using high-field magnetization and present the magnetic phase diagram of this system. Elastic neutron scattering measurements under applied magnetic fields of up to 10 T are discussed in Sec.~\ref{neutron} followed by the conclusion in Sec.~\ref{conclusion}

\section{Experiment}\label{experiment}

The single crystals of $\alpha$-Cu$_{2}$V$_{2}$O$_{7}$ studied in this paper were grown by the vertical Bridgman technique. The detailed method of crystal growth and characterization are described elsewhere~\cite{cvo1}.  The crystals with dimensions of about 4 $\times$ 4 $\times$ 4 mm$^3$ were aligned using a four-circle X-ray diffractometer with Mo$K\alpha$ radiation and cut perpendicular to the crystallographic $b$- and $c$-axes (the $a$-axis is the naturally cleaved facet). Magnetic properties at low fields (up to 7~T) were studied using a superconducting quantum interference device (MPMS-XL, Quantum Design) down to the base temperature of 1.8~K. Magnetization as a function of field and temperature was measured when the magnetic field was applied parallel to each of the crystallographic axes. To study the magnetic properties at high fields, the nondestructive pulsed magnet at the International MegaGauss Science Laboratory, Institute for Solid State Physics (ISSP), University of Tokyo was used to generate pulsed magnetic fields of up to 56 T.  Magnetization was measured by induction using a coaxial pick-up coil.  The single-crystal sample was aligned so that the applied field was either parallel or perpendicular to the $a$-axis, and cooled to the base temperature of 1.4 K using a liquid $^4$He cryostat.  

The DFT calculations were performed using the \textsc{quantum espresso} simulation package~\cite{espresso}. All calculations were done within the generalized gradient approximation in the form of Perdew, Burke and Ernzerhof (PBE) \cite{PBE} for
the exchange and correlation potentials with the hubbard $U$ correction (GGA+$U$)
in order to explicitly take into account the correlated effect of the 3d electrons of
Cu$^{2+}$ ions. We adopted the values of the on-site Coulomb and exchange interaction
parameters $U$ = 7.0 eV and $J$~=~0.5 eV according to similar compounds\cite{Tsirlin, Yashima}.   To
cross-check the choice of the Coulomb parameters, we calculated the electronic
structure e.g. a band gap for several values of $U$ and evaluated the
exchange coupling for $U$ = 7.0 eV and $U$ = 8.0 eV. The  effect of core electrons
was modeled through the use of ultrasoft pseudopotentials with the planewave
cutoff of 80 Ry. The Gaussian broadening technique was used and meshes of $2
\times 4 \times 4 $ and $4 \times 6 \times 6$ $k$-points were sampled for the
Brillouin-zone integrations.  All calculations were done with the experimental
crystal structure  whose lattice parameters are  $a=20.6786$ \AA, $b =
8.4052$ {\AA} and $c=6.4462$ \AA. \cite{cvo1}  The internal lattice
coordinates from the experimental measurements were also used in the
calculations.  The crystal structure of $\alpha$-Cu$_2$V$_2$O$_7$ belongs to
the $Fdd2$ spacegroup thus yielding the 88-atom unit cell.  To address the
consistency of the structural data, we performed the structural relaxation; the
discrepancy of the atomic coordinates is less than 0.2 {\AA} and the forces
do not exceed 0.001 Ry/a.u. This small distortion in the atomic coordinates
weakly affects the electronic structure and the exchange coupling. The obtained exchange parameters were then used to construct a spin network for the QMC simulation with \textsc{loop} algorithm~\cite{loop} using the simulation package \textsc{alps}~\cite{alps} to calculate the magnetic susceptibility for comparison with the experimental data.

Finally the spin-flop state was investigated microscopically using elastic neutron scattering at the SPINS instrument, NIST Center for Neutron Research (NCNR), USA. The single crystal of mass 1.39 g was alinged so that the $bc$-plane was in the scattering plane. The fixed final neutron energy of 5~meV was utilized with the horizontal collimations of open -- 80$'$ -- sample -- 80$'$ -- detector. The vertical magnetic field between 0 to 10~T was applied along the crystallographic $a$-axis to investigate the spin-flop transition and the magnetic structure of $\alpha$-Cu$_{2}$V$_{2}$O$_{7}$ in the spin-flop state.

\section{Results and discussion}\label{resultanddiscussion}

\subsection{Low-field magnetization}\label{mpms}

In our previous work,\cite{cvo1} the magnetization as a function of magnetic field $M(H)$ on single-crystal $\alpha$-Cu$_{2}$V$_{2}$O$_{7}$ was measured with the applied magnetic fields of up to 7 T along two orthogonal directions, i.e., $H \parallel a$ and $H \perp a$. The results showed magnetic anisotropy between the $a$-axis and $bc$-plane. Weak ferromagnetism, which suggests canted moments as a result of the DM interaction, was observed in the ordered state for $H\perp a$. A later study on this system by Lee {\it et al.}~\cite{cvo2} revealed, from the magnetization measurements along all three crystallographic axes, that the spins are only canted along the $c$-axis and the canting angle varies from 2$^\circ$ to 7$^\circ$ depending on the applied magnetic field from 0 to 9~T. As a result, the relevant DM vectors between the pairs of nearest-neighbors (${\mathbf D_{ij}\cdot S_i \times S_j}$) can only point along the $b$-axis given the collinear spin structure along the $a$-axis.  In this work, we performed a detailed investigation of the magnetization as a function of field as well as magnetic susceptibility as a function of temperature on the aligned single crystals when the magnetic field was applied along all three crystallographic axes. The samples studied in this work are from the same batch as those reported in our previous study~\cite{cvo1}. 

\begin{figure}
	\begin{center}
		\includegraphics[width=\columnwidth]{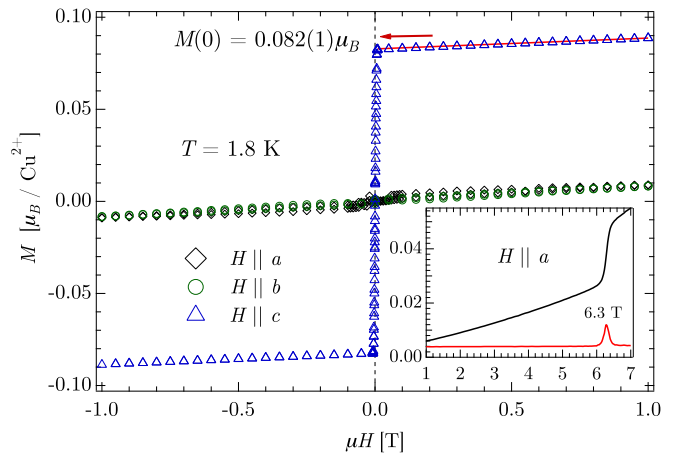}
		\caption{(Color online) Magnetization as a function of field at 1.8 K near the zero field which is applied along the $a$-axis (black diamonds), $b$-axis (green circles) and $c$-axis (blue triangles). The solid line is a linear fit to the magnetization at $H >0.1$~T and interpolated to $H \rightarrow 0$. Inset: the magnetization along the $a$-axis up to the field of 7~T shows the magnetic phase transition at 6.3 T indicated by a peak in d$M$/d$H$ and denoted by the red line.}\label{fig2}
	\end{center}
\end{figure}

Figure~\ref{fig2} shows the magnetization as a function of field between $-1$~T and 1~T for the applied field along each of the crystallographic axes at 1.8 K. These results confirm that the weak ferromagnetism exists only for the field along the $c$-axis, where the spontaneous magnetization is clearly observed, in agreement with the work by Lee {\it et al.}~\cite{cvo2}. The magnetization as the field approaches zero $M(0)$ is determined from the linear fit for $H > 0.1$ T. The interpolation gives $M(0)$ = 0.082(1)$\mu_B$, from which the canting angle $\eta$ can be calculated using $\eta=\sin^{-1}\frac{M(0)}{g\mu_B S}$ yielding $\eta=4.7(1)^\circ$. Note that the value of $M(0)$ in our previous report~\cite{cvo1} was not precisely determined since the magnetic field was applied perpendicular to the $a$-axis but not precisely along the $c$-axis.  The different values of $M(0)$ suggest that the applied field in Ref.~\onlinecite{cvo1} was $\sim 30^\circ$ away from the $c$-axis.

\begin{figure}
	\begin{center}
		\includegraphics[width=\columnwidth]{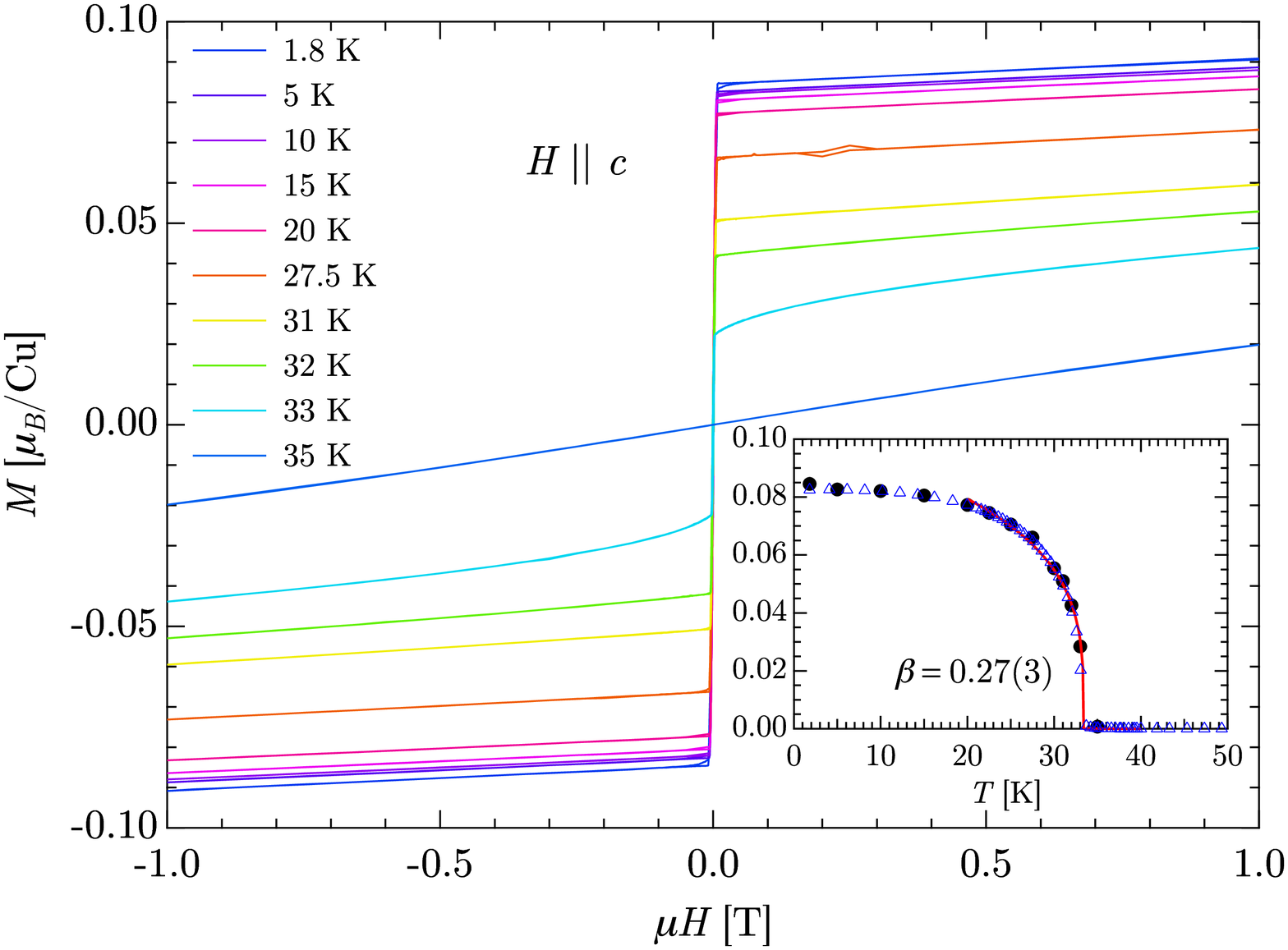}
		\caption{(Color online) Magnetization as a function of applied field with $H \parallel c$ at different temperatures from 1.8 K to 35~K (only selected temperatures are shown). Inset shows the power-law fit to the magnetization at zero field $M(0)$ (black circles). Error bars are smaller than the plot symbol. The blue triangle in the inset is the magnetization as a function of temperature when the field of $H$ = 100 Oe is applied along the $c$-axis.}\label{fig3}
	\end{center}
\end{figure}

The magnetization along the $a$- and $b$-axis, on the other hand, show a linear relation through zero field implying that the spin component along those axes are antiparallel resulting in zero net spontaneous magnetization, which is also consistent with the magnetic structure reported earlier~\cite{cvo1, cvo2}. Since the canting is along the $c$-axis and the spins anti-align along the $a$-axis, the relevant component of the DM vector, which in our previous work was proposed to lie within the $bc$-plane, must be solely along the $b$-axis. Interestingly, when the field is applied along the $a$-axis, a magnetic phase transition appears at 6.3 T as shown in the inset of Fig.~\ref{fig2}. This magnetic phase transition, which is not observed when $H \parallel b$, is due to the spin-flop transition and will be discussed in detail in Section~\ref{highfield}. 

A series of $M(H)$ measurements at different temperatures (Fig.~\ref{fig3}) shows that the remnant magnetization and hence the value of $M(0)$ decreases as temperature increases; $M(0)$ goes to zero at $T_N$ (the inset of Fig.~\ref{fig3}). A fit of the measured temperature dependence of $M(0)$ to the power-law $M(0,T)\propto\left(1-T/T_N\right)^{\beta}$ for $20~\textrm{K}<T<33.4~\textrm{K}$ yields $\beta$~=~0.27(3). This value of the critical exponent is quite close to that obtained from the order parameter measurement of the magnetic Bragg intensity using neutron scattering [$\beta$ = 0.21(1)]~\cite{cvo1}. The inset also shows the field-cooled magnetization, measured at the low-field of $100$~Oe along the $c$-axis, as a function of temperature which, as expected, perfectly follows the temperature dependence of $M(0)$.

\begin{figure}
	\begin{center}
		\includegraphics[width=\columnwidth]{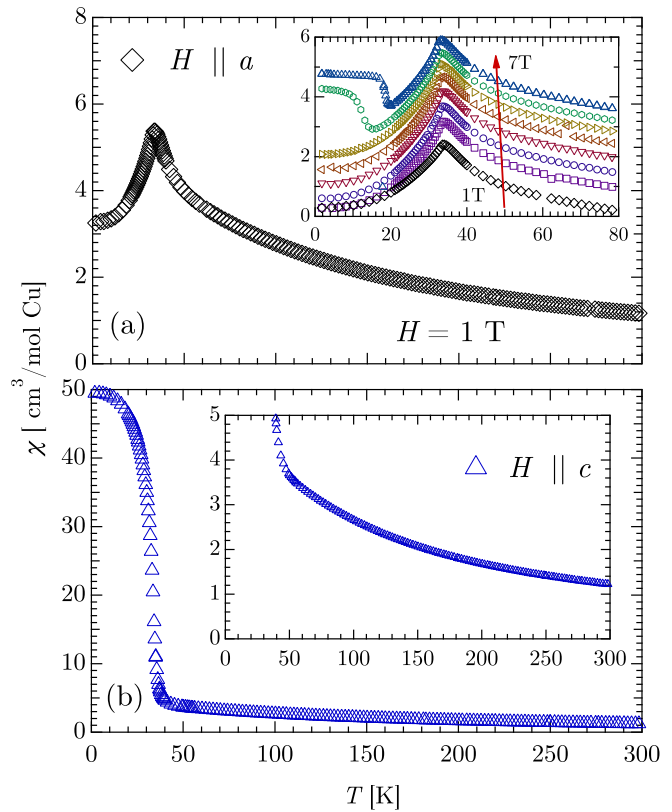}
		\caption{(Color online) Temperature dependence of the magnetic susceptibility when a field of 1~T is applied (a) along the $a$-axis and (b) $c$-axis. Inset in (a) shows the N\'eel transition at different applied field from 1~T (black diamonds) to 7~T (blue triangles) with $y$-offset. The inset in (b) shows a clear and smooth decrease in the magnetic susceptibility as the temperature increases following the Curie-Weiss law.}\label{fig4}
	\end{center}
\end{figure}

The magnetic susceptibility measured at the applied field of 1 T along the $a$- and $c$-axis are shown as a function of temperature  in Fig.~\ref{fig4}. The data for $H \parallel b$ (Fig.~\ref{fig8}) will be discussed in Section~\ref{dft}. When the field is applied along the $a$-axis, there is a sharp N\'eel transition at $T_N \approx$ 35 K which, as shown in the inset of Fig.~\ref{fig4}(a), slightly decreases toward lower temperature when the applied field is increased (see Fig.~\ref{fig11} for the $H(T)$ phase diagram). When $H \parallel c$, there is a spontaneous magnetization below $T_N$ due to the spin canting as described above. The value of the remnant magnetization as $T \rightarrow 0$ along the $c$-axis is much higher than that along the other two axes. Above 50~K the magnetic susceptibility shows a clear and smooth curve following the Curie-Weiss law up to 300 K. It should be noted that the previously observed broad peak in the magnetic susceptibility data for $H\perp a$ around $T$ = 50~K can now be observed only in the $H \parallel b$ data [Fig.~\ref{fig8}(a)].  This board peak will be analyzed and fitted in the next section. 

\subsection{Density functional theory calculation \& Quantum Monte Carlo simulation}\label{dft}

In order to derive the exchange interactions between the Cu--Cu couplings, we
performed total energy calculations for 120 different magnetic structures
including the ferromagnetic, antiferromagnetic and other spin
configurations.  The calculations show that structures with ferromagnetic and
random spin structures are more energetic than the antiferromagnetic
structure. The energy of the antiferromagnetic ordered state is about 3.3~meV
per formula unit cell  lower than the others. Therefore, it is in agreement
with the known ground state of $\alpha$-Cu$_2$V$_2$O$_7$.

\begin{figure}
	\begin{center}
		\includegraphics[width=\columnwidth]{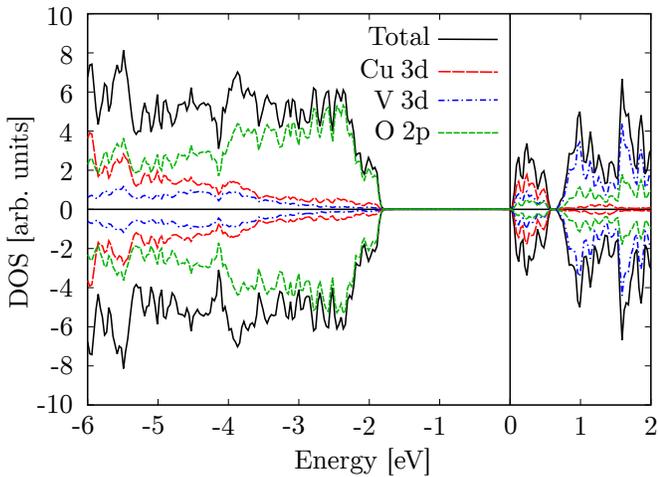}
		\caption{(Color online) Total and atomic-resolved density of states per formula unit of the $\alpha$-Cu$_2$V$_2$O$_7$ in the collinear antiferromagnetic state. The positive and negative DOS refer to the spin up and spin down contributions,
respectively. The Fermi energy is set to zero.}\label{fig5}
	\end{center}
\end{figure}

\begin{figure}
	\begin{center}
		\includegraphics[width=\columnwidth]{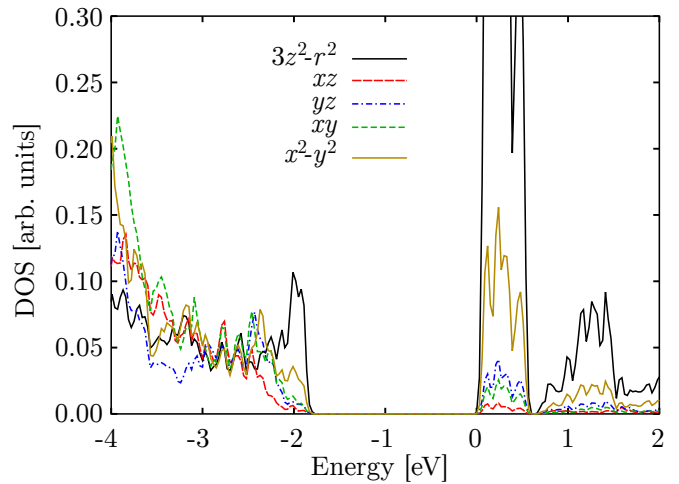}
		\caption{(Color online) Projected density of states (spin up only)  of the five Cu 3d orbitals. The Fermi level is set to zero.}\label{fig6}
	\end{center}
\end{figure}

The total and atomic-resolved density of states (DOS) of  the
ground state of $\alpha$-Cu$_2$V$_2$O$_7$ is shown in Fig.~\ref{fig5}. The Fermi level is at zero energy.
The DOS of spin-up and spin-down electrons are symmetric as expected for an
antiferromagnetic state.  The band gap is estimated to be about
1.8 eV, thus rendering the system an insulator. The bottom of the
conduction band comprises the Cu 3d, V 3d and O 2p electrons, whereas the
top of the valence band is primarily composed of the O 2p electrons with some
contributions from Cu 3d and V 3d.  It is evident that the O 2p orbitals hybridizes strongly
with the Cu 3d and V 3d orbitals in the valence band region.  To elucidate the electronic
nature and chemical bonding of the system, we plotted the orbital-resolved
density of states of the Cu 3d orbitals as depicted in Fig.~\ref{fig6}.  The
magnetic Cu$^{2+}$ ions in $\alpha$-Cu$_2$V$_2$O$_7$ have been regarded as having
a distorted octahedral environment as a result of the Jahn-Teller effect~\cite{Pommer2003, zhang}. The
d$^9$ electronic configuration of Cu$^{2+}$ implies the splitting of the
crystal field into the (t$_{2g}$)$^6$ and (e$_g$)$^3$ orbitals, which consist of the
$xy$, $xz$ and $yz$ orbitals and the $x^2-y^2$ and
$3z^2-r^2$ orbitals, respectively.  This  implies that the lower lying t$_{2g}$ orbitals are
fully filled, while the e$_g$ orbital is partially filled.  Hence, the
e$_g$ orbitals would play a crucial role for the hybridization with O 2p as
evidenced by Fig.~\ref{fig6}.     Here most of the states in
the vicinity of the Fermi energy belong to the e$_g$ contribution, i.e.,
$3z^2-r^2$ and $x^2-y^2$ with an especially large contribution from the $3z^2-r^2$ orbitals near
the Fermi energy indicating that these orbitals are magnetically active.  In contrast, the states
of the t$_{2g}$ orbitals lie in the lower energy range of $-7.5$ eV to $-4$ eV.

We evaluated the exchange interaction through  the isotropic Heisenberg model
of spin interactions whose  Hamiltonian is expressed as 
\begin{equation}
\hat{H}=\sum _{ij} J_{ij} \hat{S}_i \cdot \hat{S}_j,\label{ham}
\end{equation}
where $J_{ij}$ denotes the
coupling interaction between spins at the lattice sites $i$ and $j$.  To
quantitatively extract the coupling constant, magnetic unit cells with
different spin configurations are considered. Since the crystal structure of
$\alpha$-Cu$_2$V$_2$O$_7$ is known to have space group $Fdd2$, lower
dimensional structures can be easily utilized to define the three dominant
magnetic coupling constants, one intrachain interaction and two interchain interactions.  In the $bc$-plane, Cu$^{2+}$ cations form zigzag chains connected by two inequivalent
O$^{2+}$ ions. The coupling $J_1$ corresponds to the first nearest-neighbor Cu--Cu
with the shortest intrachain bond of 3.138 {\AA}.  Another lower-dimensional
structure linking all the 1-D chains in the crystal to form a network of
the intertwining spin-chains defines the other two coupling constants, $J_2$ and $J_3$.
The coupling $J_2$ emerges from the two Cu$^{2+}$ ions of different chains via  the
shorter 3.982~{\AA} bonds while $J_3$ relates to the longer bond of 5.264~{\AA}
as depicted in Fig.~\ref{fig1}. For each magnetic spin configurations, the pair
energy of the parallel and antiparallel alignments corresponding to each of the
coupling contants ($E_{FM, J_i}$ and $E_{AFM, J_i}$) and the total energies are
mapped to the Heisenberg model Hamiltonian. The coupling constants are then
determined by least-square fitting.  The calculated values of the exchange
interactions are $ J_1 = 3.02$~eV, $J_2 = 3.40$ eV and $J_3 = 6.12$~eV. 

\begin{figure*}
	\begin{center}
		\includegraphics[width=6in]{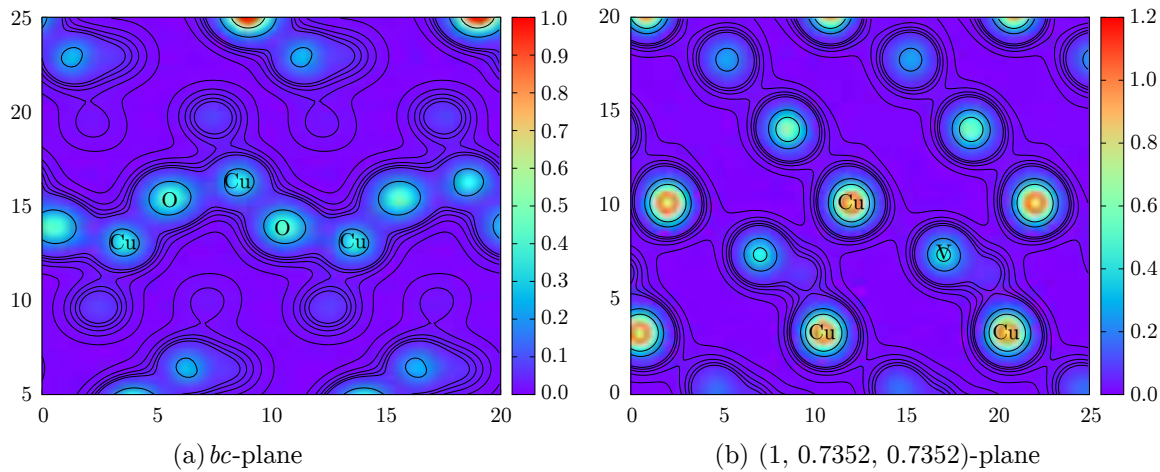}
		\caption{(Color online) Isosurface of electron density at (a) the $bc$-plane indicating 
the Cu zigzag chain and (b) the (1,~0.7352,~0.7352)  plane facilating  the $J_2$ and $J_3$ superexchange pathways.}\label{fig7}
	\end{center}
\end{figure*}

Figure~\ref{fig7} shows the isosurface of the valence electron density of
$\alpha$-Cu$_2$V$_2$O$_7$ for two different planes depicting the intrachain and
interchain coupling between the magnetic Cu$^{2+}$ ions.  Here the intrachain
Cu$-$Cu coupling can be observed through the charge density on the $bc$-plane
as shown in Fig.~\ref{fig7}(a).  Strong covalancy between Cu 3d and O 2p atomic
orbitals is observed, underlying the $J_1$ coupling.  In contrast,
Figure~\ref{fig7}(b) depicts two superexchange pathways corresponding to the two
interchain interactions. The second nearest-neighbor interaction $J_2$ is attributed to the Cu--O--Cu  pathway while the third nearest-neighbor interaction $J_3$
connects the two Cu atoms via the Cu--O--V--O--Cu pathway.  It is clear
that the charge distribution crossing the Cu--O--Cu pathway is finite but a more
pronounced distribution can be observed along the Cu--O--V--O--Cu pathway. This result indicates
that the strong exchange coupling $J_3$ is induced by the superexchange bridge
by the V d$^{5}$ orbitals. This is reasonable since the Cu--O distances in
the Cu--O--Cu pathway differ substantially (1.94 {\AA} and 3.03 {\AA}) while the Cu--O and
V--O distances in the $J_3$ coupling are comparable (ranging from 1.65 -- 1.75~\AA).  These distances are short enough to accommodate the hybridization
between the cation 3d and O 2p states. 

\begin{figure}
	\begin{center}
		\includegraphics[width=\columnwidth]{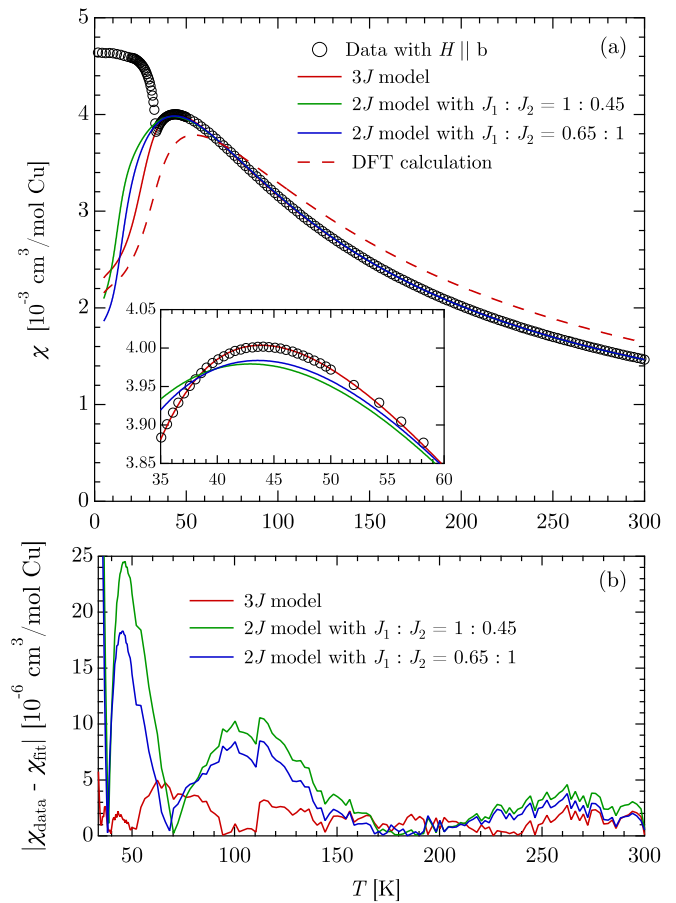}
		\caption{(Color online) Magnetic susceptibility as a function of temperature with $H \parallel b$. (a) The broad peak at around 50 K is compared to the QMC simulations with 2$J$ (green and blue line) and 3$J$ (red solid line) models. The red dashed line is a direct result from the DFT calculation. (b) The discrepancy between the calculation and data for 2$J$ and 3$J$ models.}\label{fig8}
	\end{center}
\end{figure}

The obtained values of the exchange interactions from the first-principles calculations were used to construct a spin network for the QMC simulation in order to describe the broad maximum and fit the measured magnetic susceptibility for $H \parallel b$ [Fig.~\ref{fig8}(a)]. For comparison, we used two different models; one is the 2$J$ model in which we only consider the first and second nearest-neighbor interactions $J_1$ and $J_2$, respectively, and the other is the 3$J$ model that includes the third nearest-neighbor interaction $J_3$ in the spin network (Fig.~\ref{fig1}). The values of the exchange parameters for the 2$J$ model were kept the same as those in our previous work~\cite{cvo1}, where the $J_1:J_2$ ratios of $1:0.45$ and $0.65:1$ were found to give the best fit to the experimental data for $H \perp a$. We note that the previous data is imprecise since the applied field was not perfectly aligned along the $b$-axis.  However, it is clear from our new data shown in Figs.~\ref{fig4} and \ref{fig8} that the broad peak at around 50 K only occurs when the magnetic field is applied along the $b$-axis. This broad peak is a result of short-range correlations and is related to the magnitude of the exchange couplings. To obtain a more accurate determination of the $J_i$, the magnetic susceptibility calculated from the QMC simulations were refitted to the $H \parallel b$ data. The details of the QMC simulation and fitting are described elsewhere~\cite{Johnston, cvo1}.

To re-examine our previous work, we first refitted the 2$J$ model with the same $J_1:J_2$ ratios of $1:0.45$  and $0.65:1$, the results of which are represented in Fig.~\ref{fig8}(a) by the green and blue lines, respectively. The discrepancy between the experiment and calculations especially around the broad peak shown in the residue plot of Fig.~\ref{fig8}(b) suggests that the 2$J$ model falls short of capturing the accurate spin correlations. In the inset of Fig.~\ref{fig8}(a), the maximum position of the broad peak is higher than those obtained from the calculations using the 2$J$ model, which implies that the actual average value of the $J_i$ must be higher than our previous estimation.  We then compare the data to the QMC simulation with the 3$J$ model by using the values of the $J_i$ obtained directly from the DFT calculations to construct the spin network. However, as shown by the red dashed line in Fig.~\ref{fig8}(a), the results do not fit the experimental data very well.  The discrepancy is most likely to be due to extra terms in the spin Hamiltonian, such as anisotropic exchange and antisymmetric interactions\cite{motome, swCVO}, which are not included in Eq.~\ref{ham}.  These anisotropic interactions are not fitted to the result of the DFT total energy calculations, resulting in the slight overestimate of the exchange parameters.

In order to obtain a better estimate of the $J_i$ based on the 3$J$ model, we slightly adjusted the values of exchange interactions obtained from the DFT calculations by converting them into a fraction with respect to $J_1$; this model is called the modified $3J$ model.  As a result, the $J_1:J_2:J_3$ ratio is fixed at $1:1.12:2.03$. The spin network corresponding to the three values of the exchange parameters were then used for the QMC simulation, and the calculated magnetic susceptibility was again fitted to the experimental data [red solid line in Fig.~\ref{fig8}(a)] yielding $J_1$ = 2.45(1) meV, which differs by about 20\% from the unnormalized DFT value. The fitted value of the Land\'e $g$-factor is 2.35(1), which is sufficiently close to the value of 2.44(3) obtained from the Curie-Weiss fit at high temperature ($T >$ 100~K).  The modified 3$J$ model fits the experimental data much better than the $2J$ model especially around the broad peak as shown in the inset of Fig.~\ref{fig8}(a) and in the residue plot in Fig.~\ref{fig8}(b). The obtained fitted parameters are summarized in Table~\ref{table1}.  In contrast to our previous report\cite{cvo1}, our new analysis on the broad peak at 50~K of the $H \parallel b$ data indicates that the third nearest-neighbour $J_3$ is in fact the strongest interaction, which is consistent with the previous work\cite{cvo3, cvo4}.  Using the combined DFT and QMC calculations, we were able to determine the magnitudes of the exchange interactions more accurately than before.  Furthermore, it should be noted that our DFT results for all the antiferromagnetic exchange interactions are inconsistent with the work by Sannigrahi {\sl et al.} where $J_2$ is ferromagnetic~\cite{cvo3}.  

\begin{table}
	\caption{\label{table1}Parameters obtained from the fit of magnetic susceptibility with $H \parallel b$ using different lattice models.}
	\centering
	\begin{tabular}{c c c c c}
		\hline \hline
		&~~Modified 3$J$ model~~& \multicolumn{2}{c}{2$J$ model~\cite{cvo1}}\\
		\hline
		$J_{1}$ [meV]    & 2.45(1) & ~5.84(4) &~4.25(2)\\
		$J_{2}$ [meV]    & 2.77(1) & ~2.63(2) &~6.34(3)\\
		$J_{3}$ [meV]    & 4.97(1) & ~$-$ &~$-$\\
		$g$-factor & 2.35(1) & 2.29(1) &~~2.29(1)\\
		\hline \hline
	\end{tabular}
\end{table}

\subsection{High-field magnetization}\label{highfield}

\begin{figure}
	\begin{center}
		\includegraphics[width=\columnwidth]{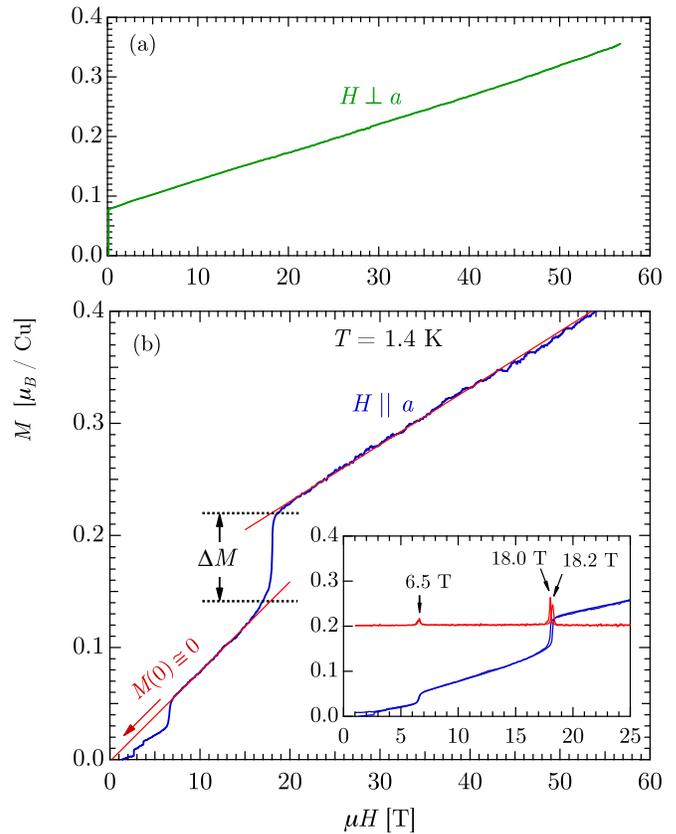}
		\caption{(Color online) Magnetization as a function of magnetic field when the field is applied parallel and perpendicular to the crystallographic $a$-axis at 1.4 K. (a) The magnetization when the field is applied perpendicular to the crystallographic $a$-axis. The main panel in (b) shows all the data up to 56~T for $H \parallel a$. The red lines are the linear fit to the data at $8~\textrm{T}<H<16~\textrm{T}$ yielding $M(0)\rightarrow$~0~T, and at $H>$ 20 T for the calculation of $\Delta M$ as described in the text. The inset shows the transition field at $H_{c1}$ = 6.5~T and $H_{c2}$ =18.2~T (18.0 T) upon the increasing (decreasing) field defined by d$M$/d$H$ in the red curve. A small amount of hysteresis can be observed at $H_{c2}$. }\label{fig9}
	\end{center}
\end{figure}

The high-field magnetization of single-crystal $\alpha$-Cu$_{2}$V$_{2}$O$_{7}$ was measured in the pulsed magnetic field applied along two orthogonal directions, i.e., $H \parallel a$ and $H \perp a$. The results at 1.4 K are shown in Fig.~\ref{fig9}. When the field is applied perpendicular to the $a$-axis [Fig.~\ref{fig9}(a)], the magnetization abruptly increases to about 0.08$\mu_B$ near zero field, which is consistent with that observed from the MPMS measurement with $H \parallel c$.  From the value of $M(0)$, it can be inferred that the $c$-axis of the crystal was closely aligned parallel to the applied field. The magnetization was found to linearly increases with the field up to 56 T without saturation or further appearance of a phase transition. On the other hand, when the field was applied along the $a$-axis [Fig.~\ref{fig9}(b)], we observed two magnetic phase transitions, indicated by the peaks in d$M$/d$H$, the first transition at $H_{c1}=6.5$~T, which was already observed in the MPMS measurement (inset of Fig.~\ref{fig2}), and the second at $H_{c2}=18.2$~T (18.0~T) upon increasing (decreasing) field.  In the ordered state, as previously stated, the $S=1/2$ Cu$^{2+}$ spins align antiparallel with their nearest and next-nearest neighbors, and the majority of the spin component is along the crystallographic $a$-axis with small field-induced canting along the $c$-axis. When the applied magnetic field along the $a$-axis is between $H_{c1}$ and $H_{c2}$ (6.5~T $< H <$ 18~T), the competition between the exchange energy and Zeeman energy forces the spins to minimize the total energy by flopping altogether into the $bc$-plane making the spin direction perpendicular to the applied magnetic field.  The majority of the in-plane spin components still approximately anti-align with their neighbors, satisfying the dominant antiferromagnetic isotropic exchange interactions. As a result, the remnant magnetization at zero field $M(0)$ in the spin-flop state also approaches zero as shown by the linear fit in Fig.~\ref{fig9}(b). In addition, as shown in the inset of Fig.~\ref{fig4}(a), the magnetic susceptibility shows only a small upturn through the spin-flop transition below $T \simeq$~20~K where the magnetic susceptibility stays constant at about 0.004$\mu_B$ as the temperature decreases toward 1.8~K. The small value of the remnant magnetization at the base temperature suggests that after the transition into the spin-flop state, the small canted moments along the $a$-axis resulting from the DM interaction remain anti-aligned as depicted in the spin diagram in Region II of Fig.~\ref{fig11}, which is consistent with the antiferromagnetic anisotropic exchange interaction in the $a$-component\cite{swCVO}. The spin-flop transition was in fact also observed in its cousin phase $\beta$-Cu$_{2}$V$_{2}$O$_{7}$ where the easy axis is along the $c$-axis~\cite{bcvo}. However, the magnetization data up to 5~T only showed a single spin-flop transition for $H \parallel c$ at around 1.5 T in contrast to the two transitions in the $\alpha$-phase.

\begin{figure}
	\begin{center}
		\includegraphics[width=\columnwidth]{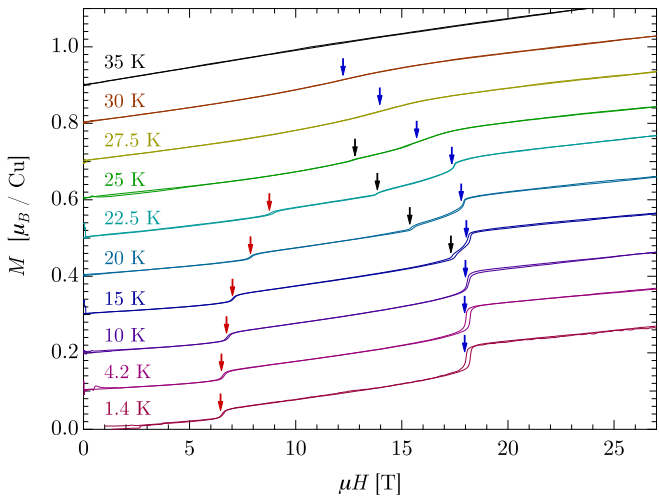}
		\caption{(Color online) Magnetization at different temperatures from 1.4 K to 35 K. The stack is due to the offset for visualization. The transition field $H_{c1}(T)$ (red arrows) and $H_{c2}(T)$ (blue arrows) denotes the spin-flop and spin-flip transitions, respectively. The third transition denoted by $H_{c3}(T)$ (black arrows) appears between T = 15 K and 25 K.}\label{fig10}
	\end{center}
\end{figure}

When the applied magnetic field reaches 18 T, we observed a second magnetic phase transition with a small hysteresis [inset of Fig.~\ref{fig9}(b)]. This second phase transition at $H_{c2} = 18$~T is a result of the Zeeman energy that overcomes the antiferromagnetic anisotropic exchange interactions making the $a$-axis component of the canted moments that previously anti-align below $H_{c2}$ align along the applied field giving rise to a non-zero $M(0)$. The change of magnetization $\Delta M$ at the antiferromagnetic-to-ferromagnetic transition at $H_{c2}$ is considerably larger than that at the spin-flop transition at $H_{c1}$.  In order to estimate the canting angle along the $a$-axis in the $H>$~18~T regime, a linear fit to the magnetization was performed to acquire the value of $\Delta M$ at $H_{c2}$, i.e., the change of magnetization where the second phase transition occurs relative to the value in the spin-flop state as depicted in Fig.~\ref{fig9}(b). The obtained high-field $\Delta M$ along the $a$-axis at 1.4 K is 0.077(1)$\mu_B$, which is slightly lower but close to the value of 0.082(1)$\mu_B$ obtained from the $H \parallel c$ data implying the same order of spin canting and a similar underlying mechanism. The value of $\Delta M$ = 0.077(1)$\mu_B$ yields a canting angle of 4.42(6)$^\circ$ along the $a$-axis. As noted above, the spin canting at low field ($H<H_{c1}$) in the $\alpha$-Cu$_{2}$V$_{2}$O$_{7}$ is due to the DM interaction at which the relevant component of the DM vector can only be along the crystallographic $b$-axis. Assuming that the direction of the DM vector does not change at high field ($H>H_{c1}$), the presence of the canted moments along the $a$-axis suggests that in the spin-flop state, it is energetically preferable for the majority of the spin component to align antiparallel along the $c$-axis.  

\begin{figure}
	\begin{center}
		\includegraphics[width=\columnwidth]{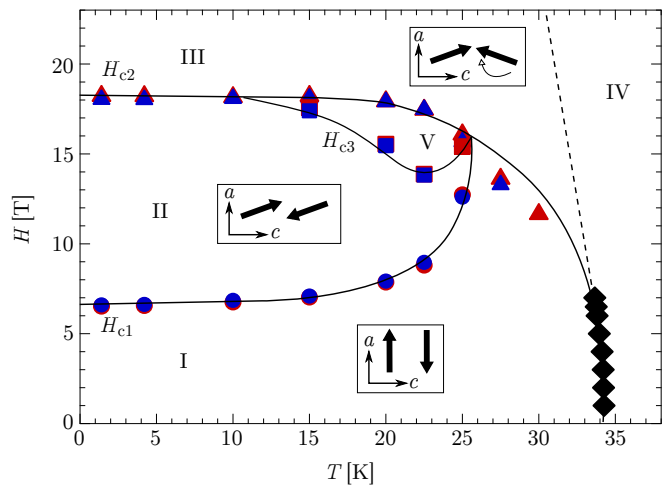}
		\caption{(Color online) Magnetic phase diagram of $\alpha$-Cu$_{2}$V$_{2}$O$_{7}$. Solid and dashed lines serve as guides to the eye. The solid lines at $H_{c1}(T)$ and $H_{c2}(T)$ represent the spin-flop and spin-flip transition, respectively, whereas $H_{c3}(T)$ represents the intermediate spin reorientation which occurs between T = 15 K and 25 K. Red (blue) symbols indicate the magnetic phase transition upon increasing (decreasing) field. The black diamond is the $H_{c2}$ obtained from the Lorentzian fit to the peak at the transition temperature of the data in the inset of Fig.~\ref{fig4}(a). The dashed line represents the cross-over between Region III and Region IV.}\label{fig11}
	\end{center}
\end{figure}

To further explore the magnetic phase transition for $H \parallel a$, the magnetization was measured at higher temperatures up to 35 K, i.e., above $T_N$. A series of data points collected from 1.4 K to 35 K is shown in Fig.~\ref{fig10}. The phase transition denoted by $H_{c1}$ and $H_{c2}$ for the first and second jumps in magnetization are indicated by the red and blue arrows, respectively. The position of $H_{c1}$ ($H_{c2}$) was found to increase (decrease) as the temperature increases toward $T_N$. The resulting critical fields as a function of temperature $H_c(T)$ are presented as a magnetic phase diagram in Fig.~\ref{fig11}. In addition, we observed the unexpected third anomaly at $H_{c3}$ as indicated by the black arrows in Fig.~\ref{fig10}, which starts to appear at $T$ = 15 K and seems to merge with $H_{c2}$ at around $T$~=~25~K. Similar behavior was also observed in the kagome lattice antiferromagnet KFe$_3$(OH)$_6$(SO$_4$)$_2$ where the spins on the alternating planes rotate 180$^\circ$ forcing the previous oppositely canted moments between the alternating layers to ferromagnetically align along the applied field~\cite{matan1}. However, it is not clear from the available data whether the same mechanism occurs in $\alpha$-Cu$_{2}$V$_{2}$O$_{7}$. We believe that there are two possible explanations for the presence of the intermediate transition at $H_{c3}$; one is the spin-rotation and the other is the spin-flip. In the former case, the applied magnetic field must simultaneously overcome both the isotropic and anisotropic interactions. On the other hand, in the latter case, it takes considerably lower energy to flip the spins along the applied magnetic field in order to overcome only the antiferromagnetic anisotropic interaction, which is much weaker than the exchange interactions.  Given that $H_{c2}=18$~T ($\sim$1~meV) at $H_{c2}$, it is most probable that the magnetic phase transition at $H_{c2}$ is due to the spin-flip and the anomaly at $H_{c3}$ is a result of the competition between the applied magnetic field and the anisotropic exchange interaction with the presence of thermal fluctuations. The dashed line in Fig.~\ref{fig11} represents the cross-over between the ordered stated in Region  III and the paramagnetic state in Region IV, which has not been resolved.  In order to verify the spin-flop state in Region II, in-field neutron scattering, which will be presented in the next section, is necessary. However, even using the strongest magnet currently available for neutron scattering, we still cannot reach the second phase transition at $H_{c2}$, making it impossible to provide further evidence for the proposed spin-flip state in Region III.

\begin{figure}
	\begin{center}
		\includegraphics[width=\columnwidth]{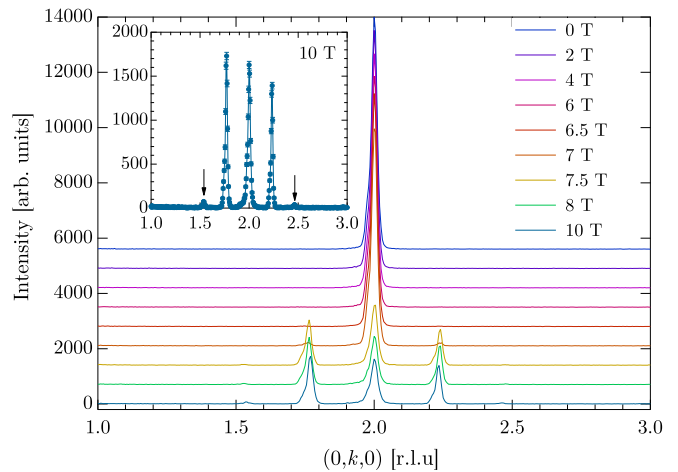}
		\caption{(Color online) Elastic neutron scattering with applied magnetic fields from 0 to 10 T along $K$ at $T$ = 2.5 K. The inset shows harmonic peaks at $H$~=~10~T, indicated by the arrows, that occurs at (0, 2$\pm2\delta$, 0) with $\delta$~=~0.23(1).}\label{fig12}
	\end{center}
\end{figure}

\begin{figure}
	\begin{center}
		\includegraphics[width=\columnwidth]{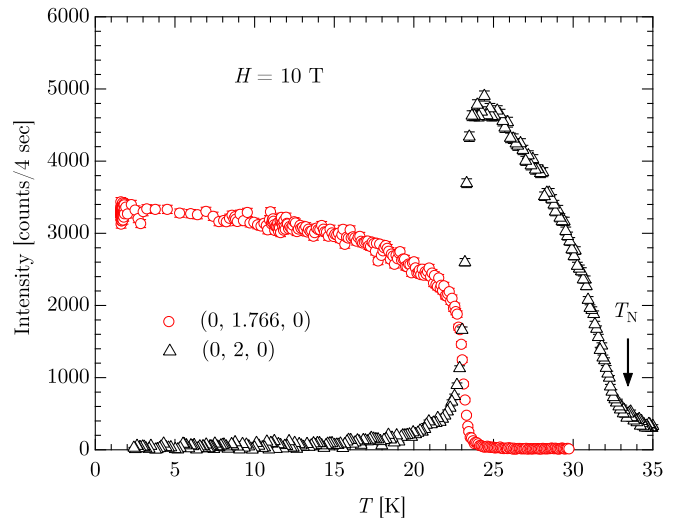}
		\caption{(Color online) Order parameter scans as a function of temperature at $H$ = 10 T of the magnetic $(0,2,0)$ (black triangles), and the incommensurate $(0,1.766,0)$ (red circle) reflections. The intensity at $(0,2,0)$ is background subtracted and divided by two. The N\'eel temperature $T_N$ = 33.4 K is indicated by the black arrow.}\label{fig13}
	\end{center}
\end{figure}

\subsection{Neutron scattering}\label{neutron}

In order to microscopically investigate the spin-flop state in Region II, elastic neutron scattering was performed on the single crystal with the applied magnetic fields of up to 10~T. The vertical field is applied along the $a$-axis with the $bc$-plane in the neutron scattering plane. The field dependence of the magnetic Bragg intensity was measured around $\mathbf{Q}=(0,2,0)$. At zero field, the spins align antiparallel along the crystallographic $a$-axis resulting in the only observable $(0,2,0)$ magnetic Bragg reflection. As the applied magnetic field is increased, the intensity of $(0,2,0)$ decreases as shown in Fig.~\ref{fig12}. On the other hand, we observed two extra Bragg peaks at $(0,{2\pm\delta},0)$ where $\delta$~=~0.23(1) for $H > 6$ T, which coincides with the first jump in the high-field magnetization data at $H_{c1}$. In addition, two much smaller Bragg peaks were observed at $\delta$~=~0.46(1) which can be interpreted as the second harmonic reflections (the arrows in the inset of Fig.~\ref{fig12}), indicative of the incommensurate magnetic structure in the spin-flop state.  The shift of the magnetic Bragg intensity from the zone center to the incommensurate wavevectors is consistent with the transition from the collinear spin structure for $H<H_{c1}$ to the in-plane helical spin structure for $H>H_{c1}$.  However, in the spin-flop state $(H_{c1}<H<H_{c2})$, we did not observe a shift of the incommensurate peaks as a function of magnetic field (Fig.~\ref{fig12}),  which suggests no change in the modulation of the helical spin structure within the spin-flop state or at least up to the field of 10 T.

The magnetic scattering intensity as a function of temperature was measured at $(0,2,0)$ and $(0,1.766,0)$ to represent the order parameters in the collinear state and spin-flop state, respectively. At 10 T, as temperature decreases from above $T_N$, the intensity of the $(0,2,0)$ magnetic Bragg reflection monotonically increases before abruptly decreasing to zero at the same temperature ($\sim23$~K) as the onset of the scattering intensity at the incommensurate $(0,1.766,0)$ reflection as shown in Fig.~\ref{fig13}. We note that the $(0,2,0)$ intensity is background subtracted and then divided by two, assuming that, at the transition from the spin-flop state to the collinear state, the two incommensurate peaks merge to form $(0,2,0)$ and their intensities combine.  However, it is clear that the maximum intensity at $(0,2,0)$ after the normalization is still higher than that at $(0,1.766,0)$. Qualitatively, this result can be explained by the fact that in the spin-flop state, the majority of the spin component lies in the $bc$-plane, i.e., the neutron scattering plane, hence resulting in a lower incommensurate magnetic intensity due to the geometric factor of the scattering intensity\cite{Shirane}.

\section{Conclusion}\label{conclusion}

We have studied the magnetic properties of single-crystal $\alpha$-Cu$_{2}$V$_{2}$O$_{7}$ by means of low-field and high-field magnetization measurements, as well as elastic neutron scattering.  The combined DFT and QMC calculations confirm that the third nearest-neighbor interaction $J_3$ is the strongest exchange coupling, in agreement with the previous studies, and refine the values of the spin Hamiltonian parameters.  The high-field magnetization measurements for $H\parallel a$ reveal two consecutive magnetic phase transitions at $H_{c1}$ and $H_{c2}$. The first transition at $H_{c1}$ is due to the typical spin-flop transition similar to that observed in its cousin phase $\beta$-Cu$_{2}$V$_{2}$O$_{7}$.  In the spin-flop state, the spins align antiparallel within the $bc$-plane with anti-aligned canted moments along the $a$-axis.  As with the previously reported canted moments along the $c$-axis, the $a$-axis canted moments are a result of the DM interaction along the $b$-axis.  The anti-alignment of the canted moments is a result of the antiferromagnetic anisotropic exchange interaction.  Neutron scattering experiments reveal that for $H_{c1}<H<H_{c2}$, the incommensurate magnetic Bragg reflections emerge suggesting the modulation of the helical magnetic structure with the majority of the spin component lying within the $bc$-plane.  The second transition at $H_{c2}$ is believed to be the spin-flip transition where the previously anti-aligned canted moments become aligned with the applied magnetic field as the Zeeman energy overcomes the anisotropic exchange energy. The magnetic phase diagram was extracted from the high-field magnetization data showing the presence of the intermediate phase, which might be related to the thermal effects, between the spin-flop and spin-flip states.  \\[3mm]

{\bf Acknowledgements}\\[1mm]

We acknowledge the support of the National Institute of Standards and Technology, U.S. Department of Commerce, in providing the neutron research facilities used in this work. The identification of any commercial product or trade name does not imply endorsement or recommendation by the National Institute of Standards and Technology. The authors would like to thank Dr.~Daisuke Okuyama, Dr. Fengjie Ma, and M. A. Allen for fruitful discussions. S.Z. and M.S. acknowledge support from the US NSF (Grant no. DMR-1409510). Work at Mahidol University was supported in part by the Thailand Research Fund Grant Number RSA5880037 and the Thailand Centre of Excellence in Physics.

\bibliography{reference}
\end{document}